\documentclass[aps,prd,nofootinbib,amsmath,amssymb,superscriptaddress,tightenlines,11pt]{revtex4}
\usepackage{graphicx}
\usepackage{dcolumn}
\usepackage{bm}
\usepackage{amssymb}
\usepackage{latexsym}
\usepackage{booktabs}
\usepackage[colorlinks, linkcolor=blue, citecolor=blue, urlcolor=blue]{hyperref}

\newcommand{\be}{\begin{equation}}
\newcommand{\ee}{\end{equation}}
\newcommand{\bq}{\begin{eqnarray}}
\newcommand{\eq}{\end{eqnarray}}

\begin{document}

\title{Holographic Ricci dark energy: Interacting model and cosmological constraints}

\author{Tian-Fu Fu}
\affiliation{Department of Physics, College of Sciences,
Northeastern University, Shenyang 110004, China}
\author{Jing-Fei Zhang}
\affiliation{Department of Physics, College of Sciences,
Northeastern University, Shenyang 110004, China}
\author{Jin-Qian Chen}
\affiliation{Department of Physics, College of Sciences,
Northeastern University, Shenyang 110004, China}
\author{Xin Zhang\footnote{Corresponding author}}
\email{zhangxin@mail.neu.edu.cn} \affiliation{Department of Physics,
College of Sciences, Northeastern University, Shenyang 110004,
China} \affiliation{Center for High Energy Physics, Peking
University, Beijing 100080, China}

\begin{abstract}
We extend the holographic Ricci dark energy model to include some
direct, non-gravitational interaction between dark energy and dark
matter. We consider three phenomenological forms for the interaction
term $Q$ in the model, namely, $Q$ is taken proportional to the
Hubble expansion rate and the energy densities of dark sectors
(taken to be $\rho_{\rm de}$, $\rho_{\rm m}$, and $\rho_{\rm
de}+\rho_{\rm m}$, respectively). We obtain a uniform analytical
solution to the three interacting models. Furthermore, we constrain
the models by using the latest observational data, including the 557
Union2 type Ia supernovae data, the cosmic microwave background
anisotropy data from the 7-yr WMAP, and the baryon acoustic
oscillation data from the SDSS. We show that in the interacting
models of the holographic Ricci dark energy, a more reasonable value
of $\Omega_{\rm m0}$ will be obtained, and the observations favor a
rather strong coupling between dark energy and dark matter.
\end{abstract}

\maketitle

Our universe is undergoing an accelerated expansion, which was
discovered via the observations of distant type Ia supernovae
(SN)~\cite{Riess98}, about a decade ago. Combined analysis of
various observational data points to a cosmological model that
contains dark energy, an exotic cosmic component with negative
pressure. What is more, dark energy is dominating the evolution of
the current universe: it occupies about $73\%$ of the total energy
of the universe. However, currently, the nature of dark energy is
still enigmatic, and so the revelation of the nature of dark energy
raises one of the biggest challenges for the modern fundamental
science.

The cosmological constant $\Lambda$, first introduced by Einstein in
1917 \cite{Einstein17}, is an important candidate for dark energy,
because it can provide a nice explanation for the accelerating
universe and can fit the observational data well. Nevertheless, the
cosmological constant is suffering from severe theoretical
challenges \cite{DErev}: one cannot explain why the theoretical
value of $\Lambda$ from the current framework of physics is greater
than the observational value by many orders of magnitude, and why
the densities of dark energy and matter are in the same order just
today while they evolve with rather different ways. Besides the
cosmological constant, there is a fairly attractive idea that the
dark energy may not be a rigescent constant but a dynamically
evolving component. In fact, many dynamical models of dark energy
have been proposed and studied in detail. Among the many dynamical
models of dark energy, the holographic model of dark energy
\cite{Li04,holoext,holofit} is very attractive since it originates
from the consideration of the holographic principle \cite{holop} of
quantum gravity. It is widely believed that the physical nature of
dark energy is in deep connection with the underlying quantum
gravity theory. Thus, the theoretical and phenomenological studies
on holographic dark energy may provide significant clues for the
bottom-up exploration of a full quantum theory of gravitation.

When considering gravity in a quantum field system, the conventional
local quantum field theory will break down due to the too many
degrees of freedom that would cause the formation of a black hole.
According to the holographic principle, one may put an energy bound
on the vacuum energy density, $\rho_{\rm vac} L^3 \leq M^2_{\rm
Pl}L$~\cite{Cohen99}, where $\rho_{\rm vac}$ is the vacuum energy
density and $M_{\rm Pl}$ is the reduced Planck mass. This bound
implies that the total energy in a spatial region with size $L$
should not exceed the mass of a black hole with the same size. The
largest size compatible with this bound is the infrared (IR) cutoff
size of this effective field theory. Evidently, the holographic
principle gives rise to a dark energy model basing on the effective
quantum field theory with a UV/IR duality. From the UV/IR
correspondence, the UV problem of dark energy can be converted into
an IR problem. A given IR scale can saturate that bound, and so one
can write the dark energy density as $\rho_{\rm de}=3c^2M_{\rm
Pl}L^{-2}$~\cite{Li04}, where $c$ is a phenomenological parameter
(dimensionless) characterizing all of the uncertainties of the
theory. Note that, hereafter, we use $\rho_{\rm de}$ to denote the
dark energy density. Now, the problem becomes how to choose an
appropriate IR cutoff for the theory. Li \cite{Li04} proposed to
choose the event horizon of the universe as the IR cutoff of the
theory. This choice generates a successful holographic model of dark
energy, explaining both the fine-tuning problem and the cosmic
coincidence problem at the same time, in some degree. Subsequently,
other holographic models of dark energy basing on the different
choices of IR cutoff were also proposed. For example, the choice of
the conformal age of the universe leads to the agegraphic dark
energy model \cite{Cai:2007us}, and the choice of the Ricci scale of
the universe gives rise to the holographic Ricci dark energy
model~\cite{Gao:2007ep}.

The present paper focuses on the holographic Ricci dark energy
(hereafter, abbreviated as RDE) model. In the RDE model, the IR
length scale $L$ is given by the average radius of the Ricci scalar
curvature ${|\cal R|}^{-1/2}$, so in this case the density of the
holographic dark energy is $\rho_{\rm de}\propto {\cal R}$. In a
spatially flat universe, the Ricci scalar of the spacetime is given
by ${\cal R}=-6(\dot{H}+2H^2)$, where $H=\dot{a}/a$ is the Hubble
expansion rate of the universe, and the dot denotes the derivative
with respect to the cosmic time $t$. Therefore, the density of dark
energy can be expressed as~\cite{Gao:2007ep}
\begin{eqnarray}
\label{rde} \rho_{\rm de}=3\alpha M_{\rm Pl}^{2}(\dot{H}+2H^{2}),
\end{eqnarray}
where $\alpha$ is a dimensionless parameter replacing $c^2$ of the
Li model \cite{Li04}. Note that, throughout the paper, we consider a
flat universe owing to the fact that the flatness of the space is an
important prediction of the inflationary cosmology and has been
confirmed by observations, e.g., the current constraint is
$\Omega_{k0}\sim 10^{-3}$ \cite{wmap7}. So, the Friedmann equation
is written as $3M_{\rm Pl}^2H^2=\rho_{\rm de}+\rho_{\rm m}$, where
$\rho_{\rm m}$ denotes the matter density. The RDE model has been
studied extensively. In Ref.~\cite{Cai:2008nk}, the holographic
meaning of RDE was revealed by investigating the causal connection
scale $R_{\rm CC}$. In Ref.~\cite{arXiv:0901.2262}, the cosmological
constraints on the RDE were studied, and the quintom feature was
found. In Ref.~\cite{arXiv:0904.0045}, it was shown that the
existence of the cosmic doomsday in the RDE model would ruin the
theoretical foundation of the scenario, and a mend from the
consideration of extra-dimension effects could erase the big-rip
singularity and leads to a de Sitter finale for the holographic
cosmos. For other relevant work, see, e.g., \cite{ricciext}.

It is natural to consider the possible interaction between dark
energy and dark matter in the RDE model. If dark energy interacts
with cold dark matter, the continuity equations for them are
\begin{eqnarray}
\label{CLL}
\dot{\rho}_{\rm de} + 3H (1+w)\rho_{\rm de} &=& -Q,\\
\label{CLM} \dot{\rho}_{\rm m} + 3H \rho_{\rm m} &=& Q,
\end{eqnarray}
where $w$ is the equation of state (EOS) parameter of dark energy,
and $Q$ is the interaction term. Note that, although $\rho_{\rm m}$
includes densities of cold dark matter and baryon matter, in this
place we use $\rho_{\rm m}$ to approximately describe dark matter
density due to the fact that the density of baryon matter is much
less than that of dark matter. In addition, owing to the lack of
mechanism of microscopic origin of the interaction, one has to
assume the forms of $Q$ phenomenologically. For interacting dark
energy model, several forms for $Q$ have been put forward
\cite{intde}. The most widely used form is $Q\propto H\rho$, where
$\rho$ denotes the energy density of the dark sectors, and usually
it has three choices, namely, $\rho=\rho_{\rm de}$, $\rho=\rho_{\rm
m}$, and $\rho=\rho_{\rm de}+\rho_{\rm m}$. Suwa and Nihei
\cite{Suwa} considered the case of $Q\propto H\rho_{\rm de}$ for the
interacting Ricci dark energy (IRDE) model. And, they placed the
cosmological constraints on this model by using the SN, cosmic
microwave background (CMB), and baryon acoustic oscillation (BAO)
data. However, it should be pointed out that, in the work of Suwa
and Nihei \cite{Suwa}, only one special case of interaction is
considered. Moreover, the observational data used in the
cosmological constraints are outdated today: 307 SN data from Union
dataset, and CMB shift parameter from WMAP 5-year observation. In
this paper, we will make some improvements. We will consider a more
general interaction term, $Q\propto H\rho$, in the IRDE model, and
place the cosmological constraints on the model by employing the
latest observational data. Specifically, the interaction term is
written as $Q=3bH\rho$, where $\rho$ denotes $\rho_{\rm de}$,
$\rho_{\rm m}$, and $\rho_{\rm de}+\rho_{\rm m}$, respectively, and
$b$ is a dimensionless coupling constant. According to our
convention, $b>0$ means that the energy transfer is from dark energy
to cold dark matter.

Combining with Eqs. (\ref{rde}) and (\ref{CLM}), the Friedmann
equation can be expressed as
\begin{eqnarray}
\label{DE} \frac{ \alpha}{2} \frac{d^{2}H^{2}}{dx^{2}} - C \frac{d
H^{2}}{dx} -D H^{2} &=&0
\end{eqnarray}
where $C=1-7\alpha/2 -3\alpha b/2$ and $D=3-6\alpha-6\alpha b$ for
$Q\propto H\rho_{\rm de}$; $C=1-7\alpha/2 +3\alpha b/2$ and
$D=6\alpha-6\alpha b+3b-3$ for $Q\propto H\rho_{\rm m}$;
$C=1-7\alpha/2$ and $D=3-6\alpha-3b$ for $Q\propto H(\rho_{\rm
de}+\rho_{\rm m})$; and $x=\ln a$ with $a$ being the scale factor of
the universe. The solution to Eq. ({\ref{DE}}) is obtained,
\begin{eqnarray}
\frac{H^{2}}{H_{0}^{2}} &=& A_{+} e^{\sigma_{+} x} + A_{-}
e^{\sigma_{-} x}, \label{solution}
\end{eqnarray}
where
\begin{eqnarray}
\label{sigmapm} \sigma_{\pm} &=& \frac{C \pm \sqrt{C^{2}+2D
\alpha}}{\alpha},
\end{eqnarray}
The initial conditions of Eq. (\ref{DE}) are
\begin{eqnarray}
\left .\frac{H^{2}}{H_{0}^{2}} \right|_{x=0} = 1,
\end{eqnarray}
and
\begin{eqnarray}
\left .\frac{1}{H_{0}^{2}}\frac{dH^{2}}{dx} \right|_{x=0} =
\frac{2}{\alpha}\Omega_{\rm de0}-4 ,
\end{eqnarray}
where $\Omega_{\rm de0}=1-\Omega_{\rm m0}$. The constants $A_{\pm}$
are given by
\begin{eqnarray}
\label{rholambda} A_{\pm} &=& \pm \frac{ 2 \Omega_{\rm de0} -\alpha
(\sigma_{\mp}+4)} {\alpha(\sigma_{+}-\sigma_{-})}.
\end{eqnarray}

In what follows, we will constrain the IRDE model by using the
latest observational data. We will use the 557 SN data from the
Union2 dataset, the CMB data from the WMAP 7-year observation, and
the BAO data from the SDSS. We will obtain the best-fitted
parameters and likelihoods by using a Markov Chain Monte Carlo
(MCMC) method.

We use the data points of the 557 Union2 SN compiled in
Ref.~\cite{14Amanullah:2010vv}. The theoretical distance modulus is
defined as
\begin{equation}
\label{eq11}
\mu_{\rm th}(z_i)\equiv5\log_{10} D_L(z_i)+\mu_0,
\end{equation}
where $z={1/a}-1$ is the redshift, $\mu_0\equiv42.38-5\log_{10} h$
with $h$ the Hubble constant $H_0$ in units of 100 km s$^{-1}$
Mpc$^{-1}$, and the Hubble-free luminosity distance
\begin{equation}
\label{eq12}
D_L(z)=(1+z)\int_0^z \frac{dz'}{E(z';{\bm
\theta})},
\end{equation}
where $E(z)\equiv H(z)/H_0$, and ${\bm\theta}$ denotes the model
parameters. Correspondingly, the $\chi^2$ function for the 557
Union2 SN data is given by
\begin{equation}
\label{eq13} \chi^2_{\rm
SN}({\bm\theta})=\sum\limits_{i=1}^{557}\frac{\left[\mu_{\rm
obs}(z_i)-\mu_{\rm th}(z_i)\right]^2}{\sigma^2(z_i)},
\end{equation}
where $\sigma$ is the corresponding $1\sigma$ error of distance
modulus for each supernova. The parameter $\mu_0$ is a nuisance
parameter and one can expand Eq.~(\ref{eq13}) as
\begin{equation}
\label{eq14}
\chi^2_{\rm SN}({\bm\theta})=A({\bm\theta})-2\mu_0
B({\bm\theta})+\mu_0^2 C,
\end{equation}
where $A({\bm\theta})$, $B({\bm\theta})$ and $C$ are defined in
Ref.~\cite{Nesseris:2005ur}. Evidently, Eq.~(\ref{eq14}) has a
minimum for $\mu_0=B/C$ at
\begin{equation}
\label{eq15}
\tilde{\chi}^2_{\rm
SN}({\bm\theta})=A({\bm\theta})-\frac{B({\bm\theta})^2}{C}.
\end{equation}
Since $\chi^2_{\rm SN,\,min}=\tilde{\chi}^2_{\rm SN,\,min}$, instead
minimizing $\chi_{\rm SN}^2$ we will minimize $\tilde{\chi}^2_{\rm
SN}$ which is independent of the nuisance parameter $\mu_0$.

Next, we consider the cosmological observational data from WMAP and
SDSS. For the WMAP data, we use the CMB shift parameter $R$; for the
SDSS data, we use the parameter $A$ of the BAO measurement. It is
widely believed that both $R$ and $A$ are nearly model-independent
and contain essential information of the full WMAP CMB and SDSS BAO
data \cite{55Wang:2006ts}. The shift parameter $R$ is given by
\cite{55Wang:2006ts,54Bond:1997wr}
 \begin{equation}\label{eq31}
   R\equiv\Omega_{\rm m0}^{1/2}\int_0^{z_\ast}\frac{d{z}}{E({z})},
 \end{equation}
where the redshift of recombination $z_\ast=1091.3$, from the WMAP
7-year data \cite{wmap7}. The shift parameter $R$ relates the
angular diameter distance to the last scattering surface, the
comoving size of the sound horizon at $z_\ast$ and the angular scale
of the first acoustic peak in the CMB power spectrum of temperature
fluctuations \cite{55Wang:2006ts,54Bond:1997wr}. The value of $R$ is
$1.725\pm0.018$, from the WMAP 7-year data \cite{wmap7}. On the
other hand, the distance parameter $A$ from the measurement of the
BAO peak in the distribution of SDSS luminous red galaxies
\cite{57Tegmark:2003ud} is given by
 \begin{equation}\label{eq32}
   A\equiv\Omega_{\rm m0}^{1/2}E(z_{\rm b})^{-1/3}\left[\frac{1}{z_{\rm b}}\int_0^{z_{\rm b}}\frac{d{z}}{E({z})}\right]^{2/3},
 \end{equation}
where $z_{\rm b}=0.35$. In Ref.~\cite{58Eisenstein:2005su}, the
value of $A$ has been determined to be $0.469\ (n_{\rm
s}/0.98)^{-0.35}\pm0.017$. Here, the scalar spectral index $n_{\rm
s}$ is taken to be $0.963$, from the WMAP 7-year data \cite{wmap7}.
So the total $\chi^2$ is given by
 \begin{equation}\label{eq33}
   \chi^2=\tilde{\chi}_{\rm SN}^2+\chi_{\rm CMB}^2+\chi_{\rm BAO}^2,
 \end{equation}
where $\tilde{\chi}_{\rm SN}^2$ is given by (\ref{eq15}), $\chi_{\rm
CMB}^2=(R-R_{\rm obs})^2/\sigma_R^2$ and $\chi_{\rm BAO}^2=(A-A_{\rm
obs})^2/\sigma_A^2$. The best-fitted model parameters are determined
by minimizing the total $\chi^2$.

\begin{table*}\caption{The fit results of the $\Lambda$CDM, RDE and IRDE models.
Here, IRDE1 model corresponds to $Q=3bH\rho_{\rm de}$, IRDE2
corresponds to $Q=3bH\rho_{\rm m}$, and IRDE3 model corresponds to
$Q=3bH(\rho_{\rm de}+\rho_{\rm m})$.}
\begin{center}
\begin{tabular}{cc   cc    cc   cc   cc   cc   cc}\hline\hline
 Model  & & $\Omega_{\rm m0}$&  & $b$ & & $\alpha$& &$\chi^{2}_{\rm min}$&& $\Delta$AIC && $\Delta$BIC\\ \hline
$\Lambda$CDM              && $0.270^{+0.014 +0.028}_{-0.013 -0.026}$
&
                   & N/A     &
                   &  N/A &
                   & 542.919 && 0 && 0 &\\
RDE               && $0.323^{+0.022 +0.037}_{-0.021 -0.035}$   &
                   & N/A     &
                   & $0.356^{+0.021 +0.035}_{-0.021 -0.035}$   &
                   & 565.683 && 24.764 && 29.090 &\\
IRDE1             && $0.273^{+0.034 +0.051}_{-0.032 -0.048}$   &
                   & $0.052^{ +0.011 +0.017}_{-0.015 -0.025}$  &
                   & $0.442^{+0.045 +0.069}_{-0.044 -0.066}$  &
                   & 542.698 && 3.779 && 12.431 &\\
IRDE2             && $0.269^{+0.034 +0.052}_{-0.032 -0.049}$  &
                   & $0.032^{+0.013 +0.021}_{-0.013 -0.019}$ &
                   & $0.433^{+0.043 +0.066}_{-0.042 -0.062}$ &
                   & 543.038 && 4.119 && 12.771 &\\
IRDE3             && $0.270^{+0.034 +0.052}_{-0.032 -0.048}$   &
                   & $0.020^{+0.006 +0.010}_{-0.007 -0.011}$  &
                   & $0.437^{+0.045 +0.068}_{-0.043 -0.064}$   &
                   & 542.875 && 3.956 && 12.608 &\\
\hline\hline
\end{tabular}
\label{table1}
\end{center}
\end{table*}

\begin{figure*}[htbp]
\begin{center}
\includegraphics[scale=0.4]{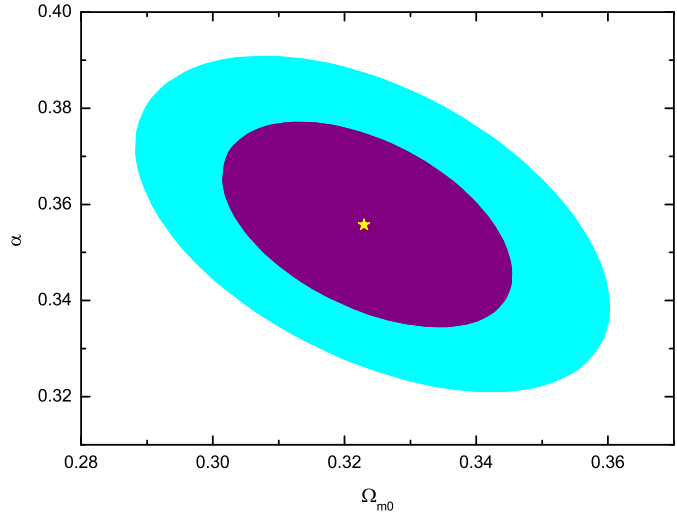}
\caption[]{\small The probability contours at $1\sigma$ and
$2\sigma$ confidence levels in the $\Omega_{\rm m0}-\alpha$ plane
for the RDE model.}\label{fig1}
\end{center}
\end{figure*}

\begin{figure*}[htbp]
\centering
\begin{center}
$\begin{array}{c@{\hspace{0.2in}}c} \multicolumn{1}{l}{\mbox{}} &
\multicolumn{1}{l}{\mbox{}}\\
\includegraphics[scale=0.32]{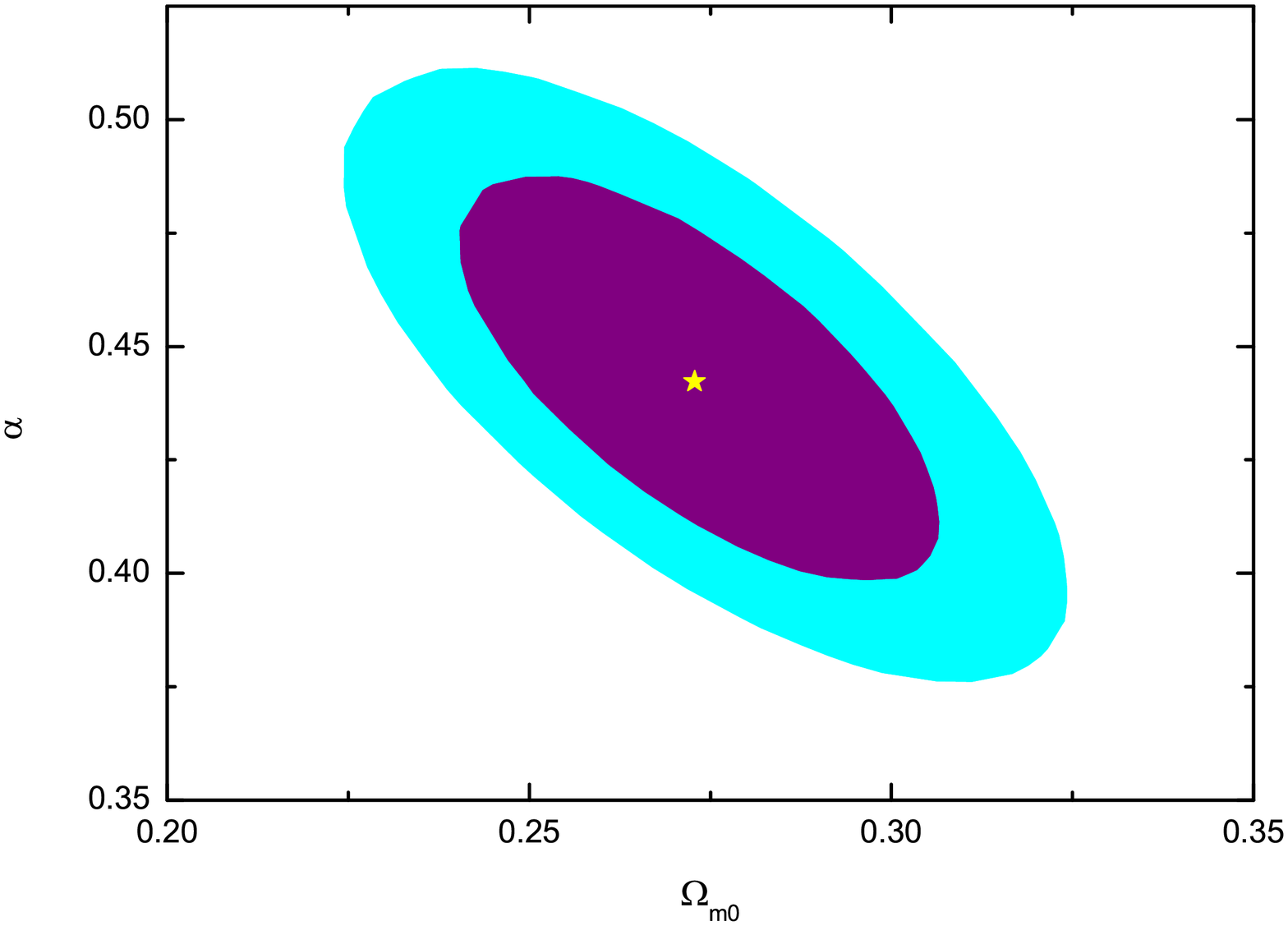} &\includegraphics[scale=0.32]{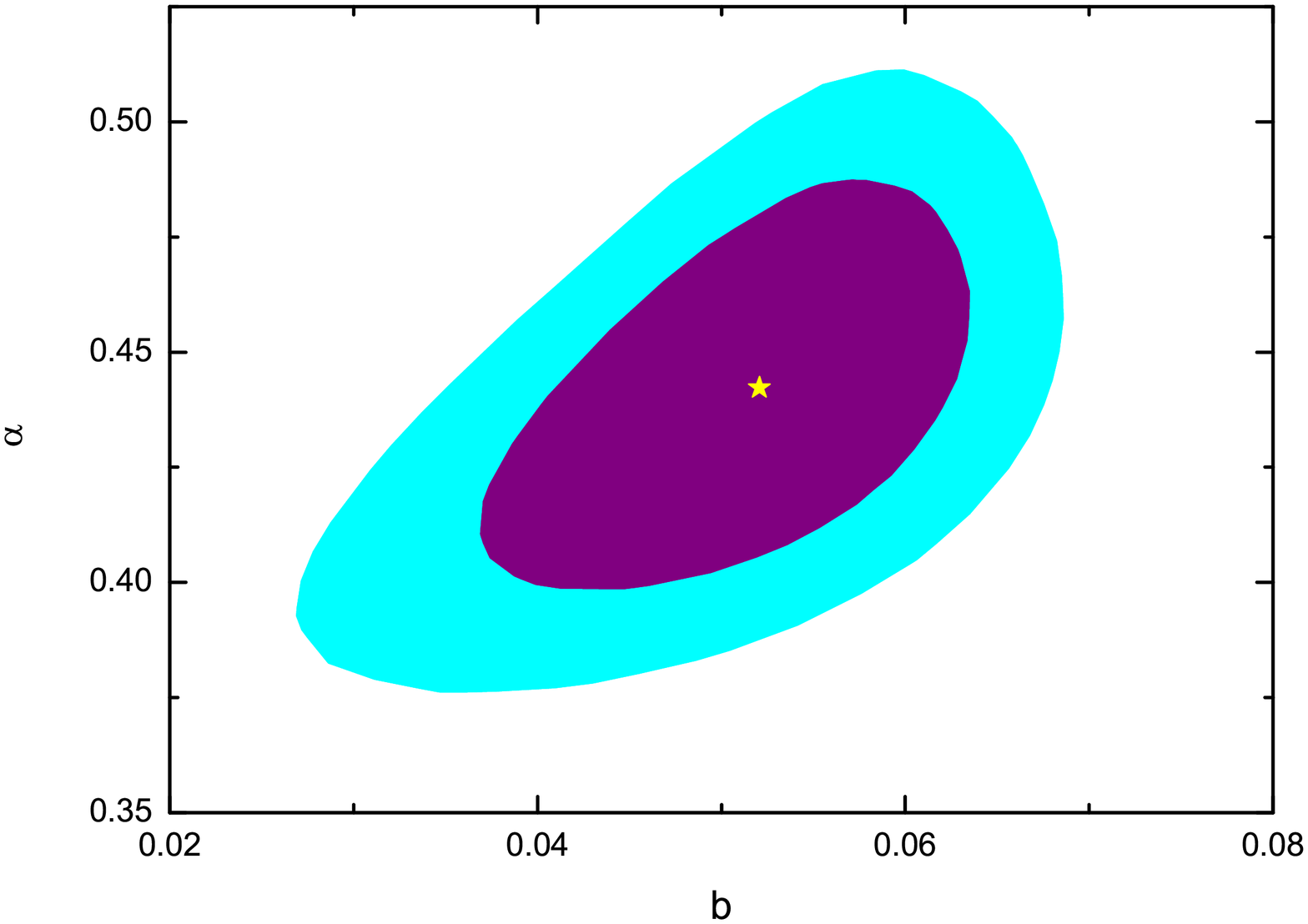}\\
\end{array}$
\end{center}
\caption[]{\small \label{fig2}The probability contours at $1\sigma$
and $2\sigma$ confidence levels in the $\Omega_{\rm m0}-\alpha$ and
$b-\alpha$ planes for the IRDE1 model (corresponding to
$Q=3bH\rho_{\rm de}$).}
\end{figure*}

\begin{figure*}[htbp]
\centering
\begin{center}
$\begin{array}{c@{\hspace{0.2in}}c} \multicolumn{1}{l}{\mbox{}} &
\multicolumn{1}{l}{\mbox{}}\\
\includegraphics[scale=0.32]{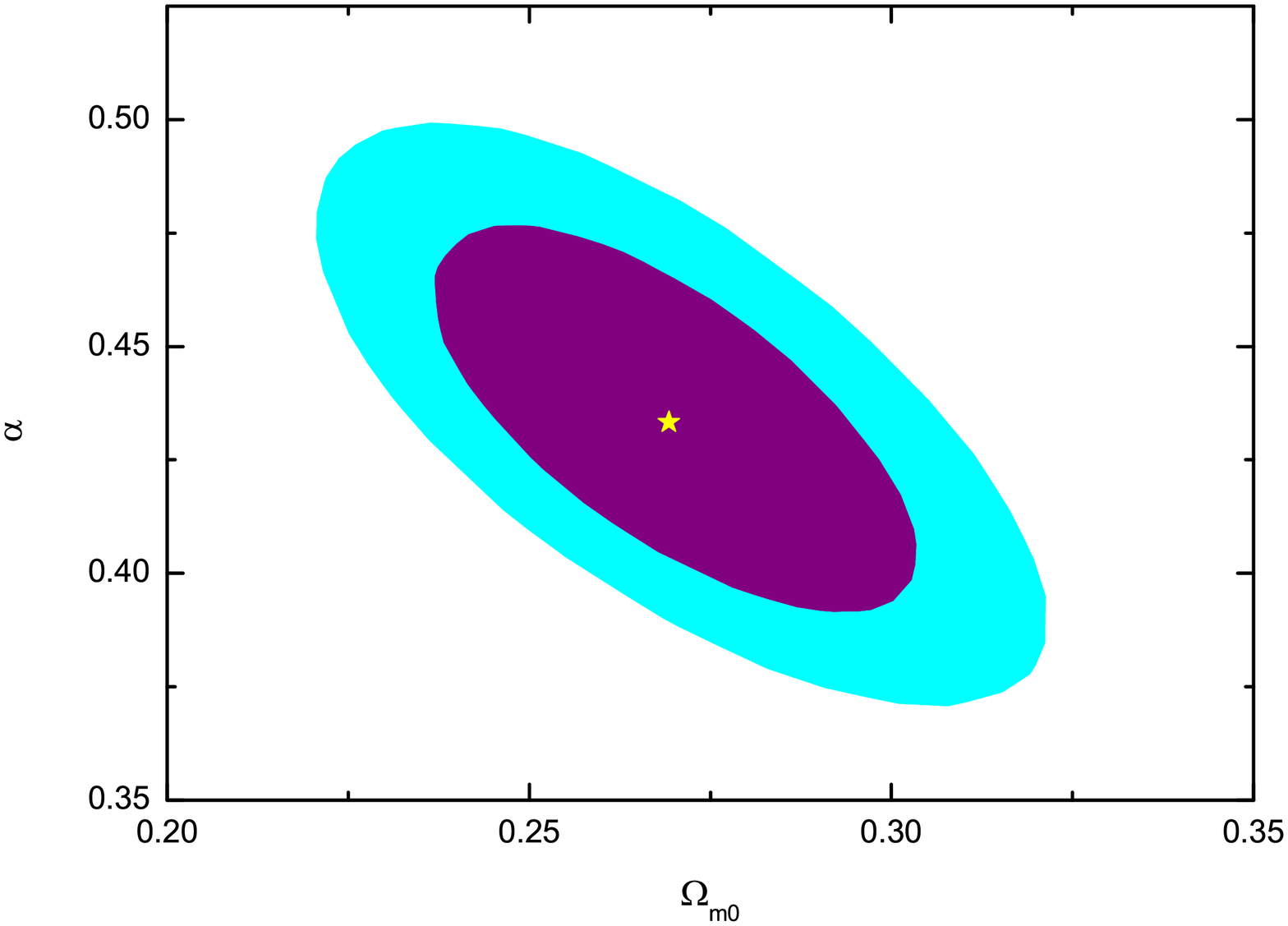} &\includegraphics[scale=0.32]{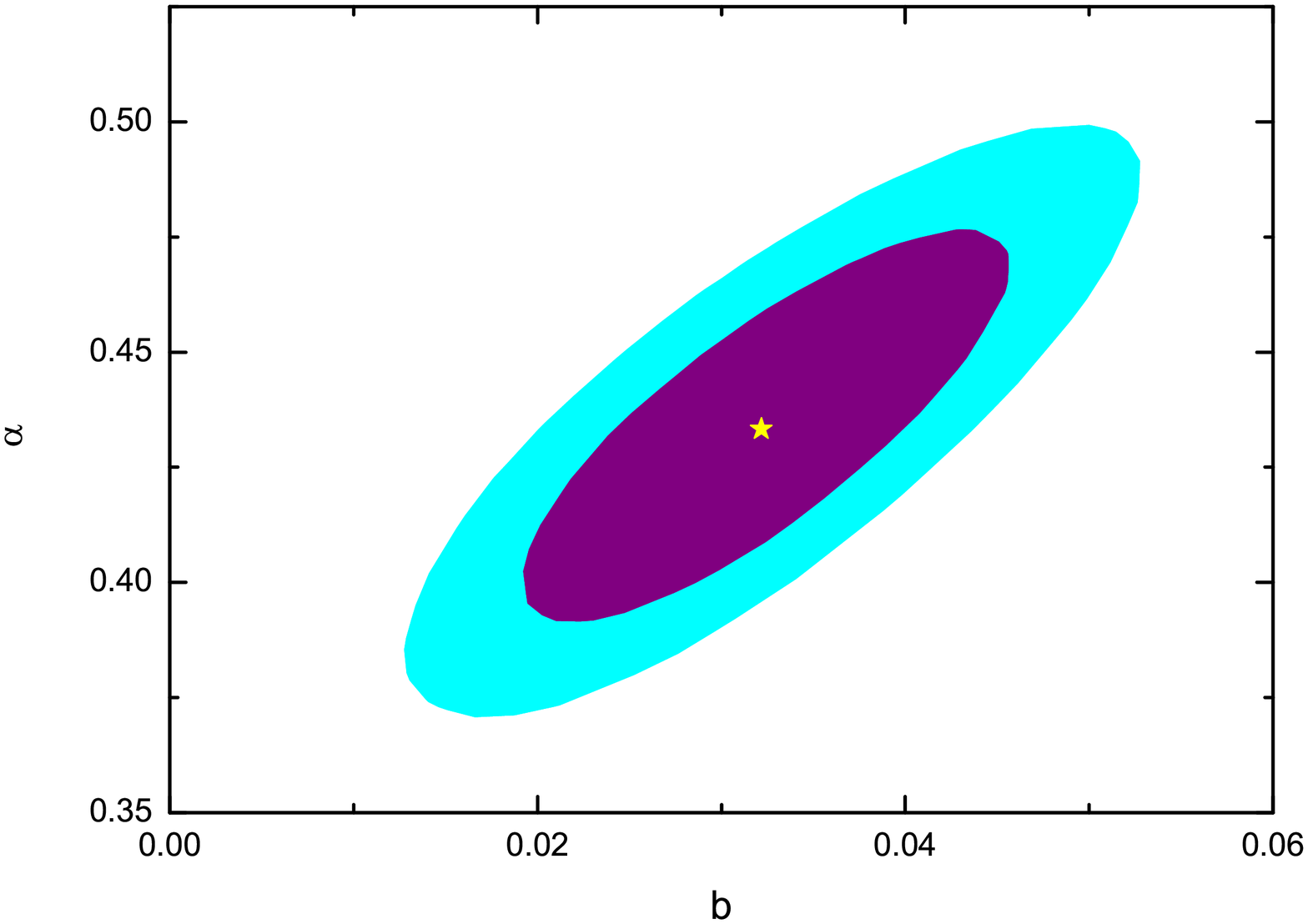}\\
\end{array}$
\end{center}
\caption[]{\small \label{fig3}The probability contours at $1\sigma$
and $2\sigma$ confidence levels in the $\Omega_{\rm m0}-\alpha$ and
$b-\alpha$ planes for the IRDE2 model (corresponding to
$Q=3bH\rho_{\rm m}$).}
\end{figure*}

\begin{figure*}[htbp]
\centering
\begin{center}
$\begin{array}{c@{\hspace{0.2in}}c} \multicolumn{1}{l}{\mbox{}} &
\multicolumn{1}{l}{\mbox{}}\\
\includegraphics[scale=0.32]{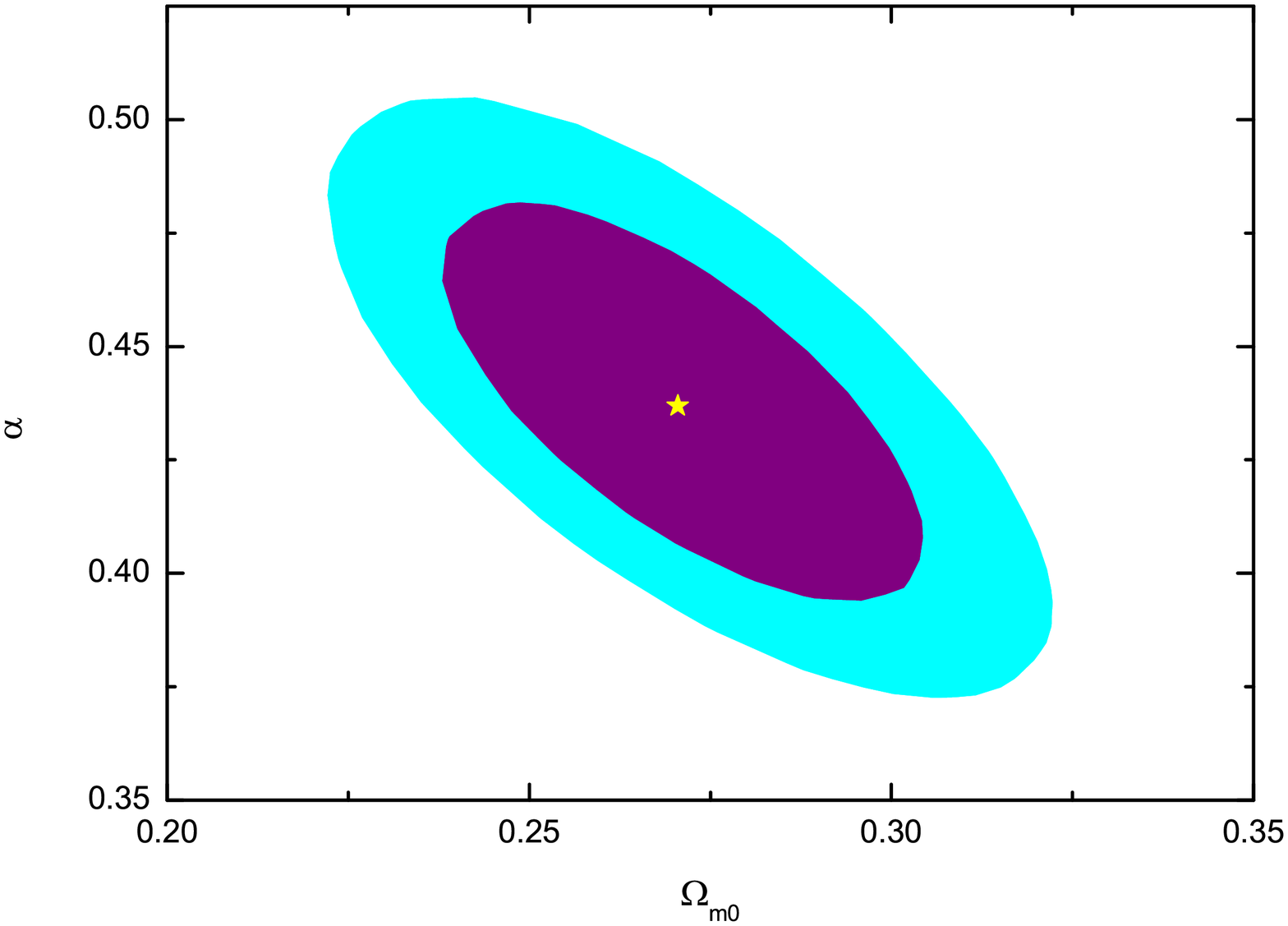} &\includegraphics[scale=0.32]{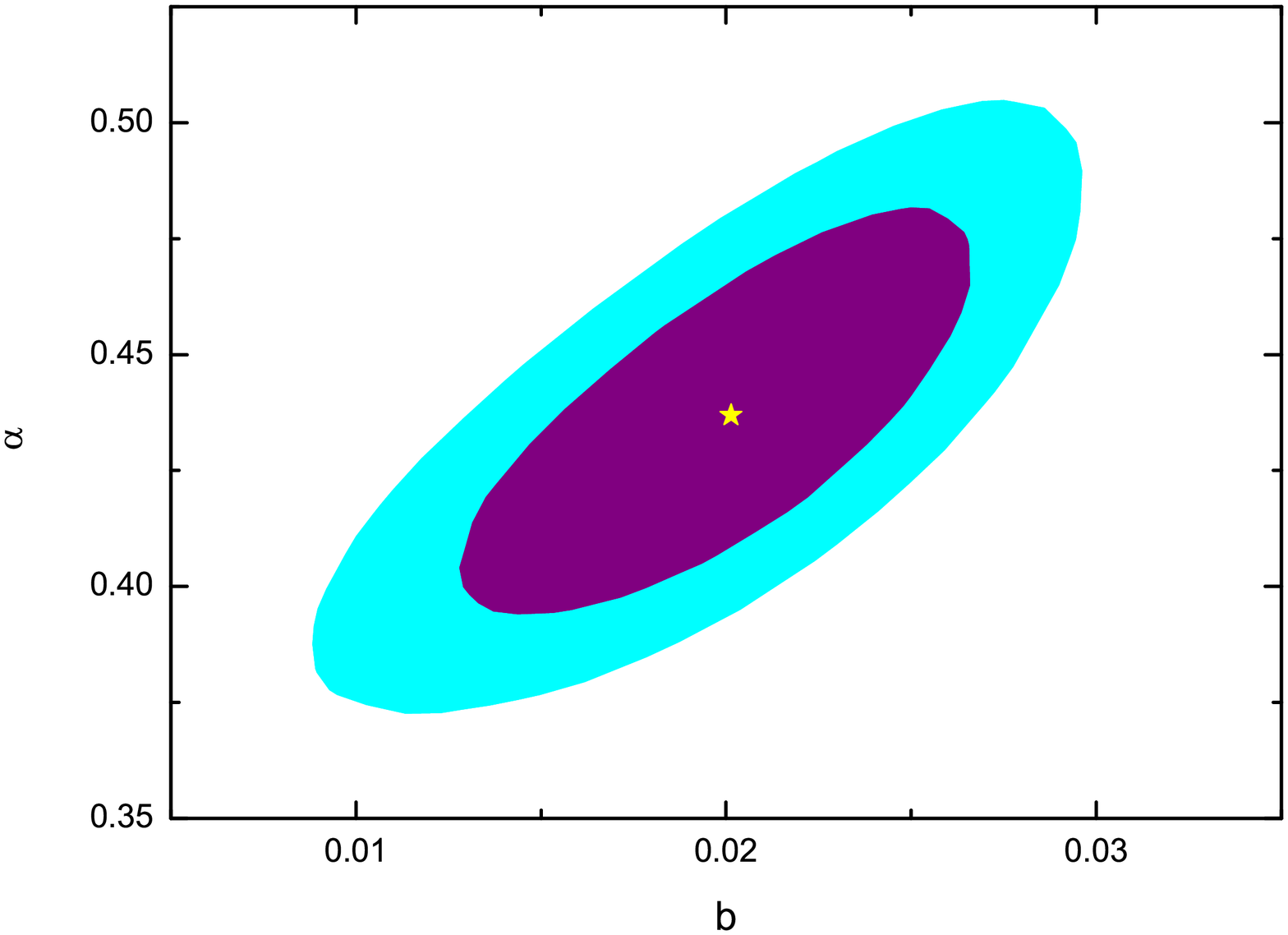}\\
\end{array}$
\end{center}
\caption[]{\small \label{fig4}The probability contours at $1\sigma$
and $2\sigma$ confidence levels in the $\Omega_{\rm m0}-\alpha$ and
$b-\alpha$ planes for the IRDE3 model (corresponding to
$Q=3bH(\rho_{\rm de}+\rho_{\rm m})$).}
\end{figure*}

Now we fit the RDE and IRDE models to the observational data. We use
the MCMC method and finally we obtain the best-fit parameters and
the corresponding $\chi_{\rm min}^2$. The best-fit, $1\sigma$ and
$2\sigma$ values of the parameters with $\chi_{\rm min}^2$ of the
four models are all presented in Table \ref{table1}. For
convenience, in the table we abbreviate the three interacting
models: IRDE1 corresponds to $Q=3bH\rho_{\rm de}$, IRDE2 corresponds
to $Q=3bH\rho_{\rm m}$, and IRDE3 corresponds to $Q=3bH(\rho_{\rm
de}+\rho_{\rm m})$.

For the RDE model, we obtain the best-fit values of the parameters:
$\Omega_{\rm m0}=0.323$ and $\alpha=0.356$, and the corresponding
minimal $\chi^2$ is $\chi_{\rm min}^2=565.683$. Obviously, the fit
value of $\Omega_{\rm m0}$ in the RDE model is rather big, evidently
greater than that of the Lambda Cold Dark Matter ($\Lambda$CDM)
model, 0.270. Note that for making a comparison with the
$\Lambda$CDM model, we also fit the $\Lambda$CDM model and list the
results also in Table \ref{table1}. Now, let us see the fit results
of the IRDE models. We find that, for all the three IRDE models, the
best-fit values obtained for $\Omega_{\rm m0}$ and $\alpha$ are
approximately equal: $\Omega_{\rm m0}=0.27$ and $\alpha=0.44$ (for
IRDE2, $\alpha=0.43$). So, it is clearly seen that the fit value of
$\Omega_{\rm m0}$ in the IRDE models is more reasonable, consistent
with that of the $\Lambda$CDM model. Comparing the values of
$\alpha$, it is found that, when there is an interaction between
dark energy and cold dark matter, the value of $\alpha$ tends to
become bigger. From the above analysis on the fitting results, we
find that the parameters $\Omega_{\rm m0}$ and $\alpha$ should be in
anti-correlation. For the coupling constant $b$ in the IRDE models,
we see clearly that the values of $b$ are explicitly different for
the three IRDE models. The value of $b$ in the IRDE1 model is the
biggest, $b=0.052$; the value of $b$ in the IRDE3 model is the
smallest: $b=0.020$. Since in the IRDE models there is an additional
parameter, $b$, the values of $\chi_{\rm min}^2$ should be
substantially less than that of the RDE model. This is the case: the
three IRDE models have the similar values of $\chi_{\rm min}^2$, all
around 543, much smaller than that of the RDE model. Among the three
IRDE model, the IRDE1 model has the smallest $\chi_{\rm min}^2$,
542.698.

However, it is unwise to use the $\chi^2$ statistic solely to
compare competing models since the number of parameters is different
for the models. In general, a model with more parameters tends to
give a lower $\chi^2_{\rm min}$, and so one may employ the
information criteria (IC) to assess the worth of models. These
statistics favor models that give a good fit with fewer parameters.
Thus, we shall use the AIC (Akaike information criterion) and BIC
(Bayesian information criterion) as model selection criteria,
defined, respectively, as
\begin{equation}\label{ic}
{\rm AIC}=\chi^2_{\rm min}+2k,~~~{\rm BIC}=\chi^2_{\rm min}+k\ln N,
\end{equation}
where $k$ is the number of parameters, and $N$ is the data points
used in the fit. Note that the absolute value of the criterion is
not of interest, only the relative value between different models,
$\Delta{\rm AIC}$ or $\Delta{\rm BIC}$, is useful. For the details
about the AIC and BIC, especially their applications in a
cosmological context, see, e.g., Refs.~\cite{IC1,IC2,IC3,IC4,IC5}.
We give the calculation results of AIC and BIC for the models in
Table~\ref{table1}. Since the $\Lambda$CDM model is an important
reference model for the studies of dark energy models, we also
include its results in the table. From Table~\ref{table1}, we see
that, though the value of $\chi^2_{\rm min}$ of the $\Lambda$CDM
model is greater than that of the IRDE models, the $\Lambda$CDM
model gives the lowest values of AIC and BIC. This result is
consistent with the previous studies \cite{IC3,IC4,IC5}. Actually,
the $\Lambda$CDM model is the best model fitting the current data
among all of the existing dark energy models. Therefore, the
$\Lambda$CDM model is chosen to be a fiducial model when we present
the values of $\Delta{\rm AIC}$ and $\Delta{\rm BIC}$. Obviously,
the RDE model is strongly unsupported by the data, since its
$\Delta$BIC is fairly high; see also Ref.~\cite{IC4} for comparison
of various dark energy models including the RDE model. Comparing
with the RDE model, the IRDE models are much better, though they are
still worse than the $\Lambda$CDM model.

Figures \ref{fig1}$-$\ref{fig4} show the probability contours at
$68.3\%$ and $95.4\%$ confidence levels in $\Omega_{\rm m0}-\alpha$
and $b-\alpha$ parameter planes for the RDE and IRDE models. We find
that the $\Omega_{\rm m0}-\alpha$ planes of the IRDE models show a
stronger degeneracy than that of the RDE model. In addition, for all
the three IRDE models, we see that $\Omega_{\rm m0}$ and $\alpha$
are anti-correlated, and $b$ and $\alpha$ are in positive
correlation.

It should also be noted that, in the work of Suwa and Nihei
\cite{Suwa}, the contour plots in the parameter planes are
incomplete: when plotting the $\alpha-\Omega_{\rm de0}$ contours,
$\gamma=3b$ is fixed to be 0.15; when plotting the $\gamma-\alpha$
contours, $\Omega_{\rm de0}$ is fixed to be 0.73; see Figs.~2 and 3
in Ref.~\cite{Suwa}. In our work, the probe of the parameter-space
is complete, i.e., when plotting the contours in parameter plane,
other nuisance parameters are marginalized. In addition, it should
be mentioned that the IRDE1 case was also investigated in
Ref.~\cite{IC5}. In Ref.~\cite{IC5}, the authors analyzed in detail
the various holographic dark energy models, with the IR cutoff
chosen to be the Hubble scale, the future event horizon, and the
Ricci scale, respectively, with an interaction between dark energy
and dark matter. For the IRDE1 model, they also constrain the model
by using the current data, but they assume $w_0=-1$ in their
analysis. The result of the coupling constant they obtained is
$b=0.05\pm 0.01$, almost the same as ours. Therefore, we find that
our study confirms the results of Ref.~\cite{IC5}. For other
relevant work, see also, e.g., Ref.~\cite{Duran:2010ky}.

In summary, we have extended the RDE model to include the possible
interaction between dark energy and cold dark matter in this paper.
Though in Refs.~\cite{Suwa,IC5} a specific form of the interaction
in the RDE model has been investigated, our study makes some
significant improvements. In this work, we have considered three
phenomenological forms for $Q$ in the RDE model, namely,
$Q=3bH\rho$, where $\rho$ can be taken to be $\rho_{\rm de}$,
$\rho_{\rm m}$ and $\rho_{\rm de}+\rho_{\rm m}$, respectively. We
have solved the IRDE models, and got a uniform analytical solution
for the three IRDE models. Furthermore, we constrained the RDE and
IRDE models by using the latest observational data, including the
557 Union2 SN data, the CMB WMAP 7-yr data, and the BAO SDSS data.
Our fit results show that when the interaction between dark energy
and dark matter is considered in the RDE model, a more reasonable
value of $\Omega_{\rm m0}$ will be obtained, i.e., $\Omega_{\rm
m0}=0.27$, rather than the value of $\Omega_{\rm m0}=0.32$ in the
RDE model without interaction. Moreover, it has been shown that a
rather strong coupling between dark energy and cold dark matter in
the RDE model is favored by the observations: the best-fit value of
$b$ is around ${\cal O}(10^{-2})$, in particular, for the IRDE1
model, $b=0.052$. Our work makes the study of the interacting model
of the holographic Ricci dark energy more complete.

\begin{acknowledgments}
This work was supported by the National Science Foundation of China
under Grant Nos.~10705041, 10975032, 11047112 and 11175042, and by
the National Ministry of Education of China under Grant
Nos.~NCET-09-0276 and N100505001.
\end{acknowledgments}



\begin{thebibliography}{99}

\bibitem{Riess98}
A.~G.~Riess {\it et al.}  [Supernova Search Team Collaboration],
  Astron.\ J.\  {\bf 116} (1998) 1009
  [arXiv:astro-ph/9805201];
S.~Perlmutter {\it et al.}  [Supernova Cosmology Project
Collaboration],
  Astrophys.\ J.\  {\bf 517} (1999) 565
  [arXiv:astro-ph/9812133].

\bibitem{Einstein17}
A.~Einstein,
  Sitzungsber.\ Preuss.\ Akad.\ Wiss.\ Berlin (Math.\ Phys.) {\bf 1917} (1917) 142.

\bibitem{DErev}
S.~Weinberg,
  Rev.\ Mod.\ Phys.\  {\bf 61} (1989) 1;
V.~Sahni and A.~A.~Starobinsky,
  Int.\ J.\ Mod.\ Phys.\  D {\bf 9} (2000) 373
  [arXiv:astro-ph/9904398];
S.~M.~Carroll,
  Living Rev.\ Rel.\  {\bf 4} (2001) 1
  [arXiv:astro-ph/0004075];
P.~J.~E.~Peebles and B.~Ratra,
  Rev.\ Mod.\ Phys.\  {\bf 75} (2003) 559
  [arXiv:astro-ph/0207347];
T.~Padmanabhan,
  Phys.\ Rept.\  {\bf 380} (2003) 235
  [arXiv:hep-th/0212290];
E.~J.~Copeland, M.~Sami and S.~Tsujikawa,
  Int.\ J.\ Mod.\ Phys.\  D {\bf 15} (2006) 1753
  [arXiv:hep-th/0603057];
  M.~Li, X.~D.~Li, S.~Wang and Y.~Wang,
  Commun.\ Theor.\ Phys.\ \ {\bf 56} (2011) 525
  [arXiv:1103.5870 [astro-ph.CO]];
  Y.~F.~Cai, E.~N.~Saridakis, M.~R.~Setare and J.~Q.~Xia,
  Phys.\ Rept.\  {\bf 493} (2010) 1
  [arXiv:0909.2776 [hep-th]];
  S.~Nojiri and S.~D.~Odintsov,
  Phys.\ Rept.\  {\bf 505} (2011) 59
  [arXiv:1011.0544 [gr-qc]].

\bibitem{Li04}
M.~Li,
  Phys.\ Lett.\  B {\bf 603} (2004) 1
  [arXiv:hep-th/0403127].

\bibitem{holoext}
  Q.~G.~Huang and M.~Li,
  JCAP {\bf 0408}, 013 (2004)
  [arXiv:astro-ph/0404229];
  Q.~G.~Huang and M.~Li,
  JCAP {\bf 0503}, 001 (2005)
  [arXiv:hep-th/0410095];
  X.~Zhang,
  Int.\ J.\ Mod.\ Phys.\  D {\bf 14}, 1597 (2005)
  [arXiv:astro-ph/0504586];
  J.~Zhang, X.~Zhang and H.~Liu,
  Eur.\ Phys.\ J.\  C {\bf 52}, 693 (2007)
  [arXiv:0708.3121 [hep-th]];
  Y.~G.~Gong,
  Phys.\ Rev.\  D {\bf 70}, 064029 (2004)
  [arXiv:hep-th/0404030];
  B.~Wang, E.~Abdalla and R.~K.~Su,
  Phys.\ Lett.\  B {\bf 611}, 21 (2005)
  [arXiv:hep-th/0404057];
  X.~Wu, R.~G.~Cai and Z.~H.~Zhu,
  Phys.\ Rev.\  D {\bf 77}, 043502 (2008)
  [arXiv:0712.3604 [astro-ph]];
  B.~Chen, M.~Li and Y.~Wang,
  Nucl.\ Phys.\  B {\bf 774}, 256 (2007)
  [arXiv:astro-ph/0611623];
  M.~Li, C.~Lin and Y.~Wang,
  JCAP {\bf 0805}, 023 (2008)
  [arXiv:0801.1407 [astro-ph]];
  M.~Li, X.~D.~Li, C.~Lin and Y.~Wang,
  Commun. Theor. Phys. {\bf 51}, 181 (2009)
  [arXiv:0811.3332 [hep-th]];
  S.~Nojiri and S.~D.~Odintsov,
  Gen.\ Rel.\ Grav.\  {\bf 38}, 1285 (2006)
  [arXiv:hep-th/0506212];
  X.~Zhang,
  Phys.\ Lett.\ B\ {\bf 683} (2010) 81
  [arXiv:0909.4940 [gr-qc]].

\bibitem{holofit}
  X.~Zhang and F.~Q.~Wu,
  Phys.\ Rev.\  D {\bf 72}, 043524 (2005)
  [arXiv:astro-ph/0506310];
  X.~Zhang and F.~Q.~Wu,
  Phys.\ Rev.\  D {\bf 76}, 023502 (2007)
  [arXiv:astro-ph/0701405];
  Q.~G.~Huang and Y.~G.~Gong,
  JCAP {\bf 0408}, 006 (2004)
  [arXiv:astro-ph/0403590];
  Z.~Chang, F.~Q.~Wu and X.~Zhang,
  Phys.\ Lett.\  B {\bf 633}, 14 (2006)
  [arXiv:astro-ph/0509531];
  J.~Y.~Shen, B.~Wang, E.~Abdalla and R.~K.~Su,
  Phys.\ Lett.\  B {\bf 609}, 200 (2005)
  [arXiv:hep-th/0412227];
  Z.~L.~Yi and T.~J.~Zhang,
  Mod.\ Phys.\ Lett.\  A {\bf 22}, 41 (2007)
  [arXiv:astro-ph/0605596];
  Y.~Z.~Ma, Y.~Gong and X. Chen,
  Eur.\ Phys.\ J.\  C {\bf 60}, 303 (2009)
  [arXiv:0711.1641 [astro-ph]];
  M.~Li, X.~-D.~Li, S.~Wang and X.~Zhang,
  JCAP\ {\bf 0906} (2009) 036
  [arXiv:0904.0928 [astro-ph.CO]];
  M.~Li, X.~-D.~Li, S.~Wang, Y.~Wang and X.~Zhang,
  JCAP\ {\bf 0912} (2009) 014
  [arXiv:0910.3855 [astro-ph.CO]].

\bibitem{holop}
  G.~'t Hooft,
  arXiv:gr-qc/9310026;
L.~Susskind,
  J.\ Math.\ Phys.\  {\bf 36} (1995) 6377
  [arXiv:hep-th/9409089].

\bibitem{Cohen99}
A.~G.~Cohen, D.~B.~Kaplan and A.~E.~Nelson,
  Phys.\ Rev.\ Lett.\  {\bf 82} (1999) 4971
  [arXiv:hep-th/9803132].

\bibitem{Cai:2007us}
  R.~G.~Cai,
  Phys.\ Lett.\  B {\bf 657}, 228 (2007)
  [arXiv:0707.4049 [hep-th]];
  H.~Wei and R.~G.~Cai,
  Phys.\ Lett.\  B {\bf 660}, 113 (2008)
  [arXiv:0708.0884 [astro-ph]];
  J.~Zhang, X.~Zhang and H.~Liu,
  Eur.\ Phys.\ J.\  C {\bf 54}, 303 (2008)
  [arXiv:0801.2809 [astro-ph]];
  X.~-L.~Liu, J.~Zhang and X.~Zhang,
  Phys.\ Lett.\ B\ {\bf 689} (2010) 139
  [arXiv:1005.2466 [gr-qc]].


\bibitem{Gao:2007ep}
  C.~Gao, F. Q. Wu, X.~Chen and Y.~G.~Shen,
  Phys. Rev. D {\bf 79}, 043511 (2009)
  [arXiv:0712.1394 [astro-ph]].

\bibitem{wmap7}
  E.~Komatsu {\it et al.} [WMAP Collaboration],
  Astrophys.\ J.\ Suppl.\ \ {\bf 192} (2011) 18
  [arXiv:1001.4538 [astro-ph.CO]].

\bibitem{Cai:2008nk}
  R.~G.~Cai, B.~Hu and Y.~Zhang,
  Commun. Theor. Phys. {\bf 51}, 954 (2009)
  [arXiv:0812.4504 [hep-th]].

\bibitem{arXiv:0901.2262}
  X.~Zhang,
  Phys.\ Rev.\ D\ {\bf 79} (2009) 103509
  [arXiv:0901.2262 [astro-ph.CO]].

\bibitem{arXiv:0904.0045}
  C.~-J.~Feng and X.~Zhang,
  Phys.\ Lett.\ B\ {\bf 680} (2009) 399
  [arXiv:0904.0045 [gr-qc]].

\bibitem{ricciext}
  C.~J.~Feng,
  arXiv:0806.0673 [hep-th];
  C.~J.~Feng,
  Phys.\ Lett.\  B {\bf 672}, 94 (2009)
  [arXiv:0810.2594 [hep-th]];
  L.~N.~Granda and A.~Oliveros,
  Phys.\ Lett.\  B {\bf 669}, 275 (2008)
  [arXiv:0810.3149 [gr-qc]];
  L.~Xu, W.~Li and J.~Lu,
  arXiv:0810.4730 [astro-ph];
  C.~J.~Feng,
  arXiv:0812.2067 [hep-th];
  K.~Y.~Kim, H.~W.~Lee and Y.~S.~Myung,
  arXiv:0812.4098 [gr-qc];
  J.~Zhang, L.~Zhang and X.~Zhang,
  Phys.\ Lett.\ B\ {\bf 691} (2010) 11
  [arXiv:1006.1738 [astro-ph.CO]].

\bibitem{intde}
  L.~Amendola,
  Phys.\ Rev.\  D {\bf 62}, 043511 (2000)
  [arXiv:astro-ph/9908023];
  D.~Comelli, M.~Pietroni and A.~Riotto,
  Phys.\ Lett.\  B {\bf 571}, 115 (2003)
  [arXiv:hep-ph/0302080];
  X.~Zhang,
  Mod.\ Phys.\ Lett.\  A {\bf 20}, 2575 (2005)
  [arXiv:astro-ph/0503072];
  R.~G.~Cai and A.~Wang,
  JCAP {\bf 0503}, 002 (2005)
  [arXiv:hep-th/0411025];
  S.~Capozziello, S.~Nojiri and S.~D.~Odintsov,
  Phys.\ Lett.\  B {\bf 632}, 597 (2006)
  [arXiv:hep-th/0507182];
  S.~Nojiri and S.~D.~Odintsov,
  Phys.\ Rev.\  D {\bf 72}, 023003 (2005)
  [arXiv:hep-th/0505215];
  J.~H.~He, B.~Wang and Y.~P.~Jing,
  JCAP {\bf 0907}, 030 (2009)
  [arXiv:0902.0660 [gr-qc]];
  J.~H.~He, B.~Wang and P.~Zhang,
  Phys.\ Rev.\  D {\bf 80}, 063530 (2009)
  [arXiv:0906.0677 [gr-qc]];
  C.~G.~Boehmer, G.~Caldera-Cabral, R.~Lazkoz and R.~Maartens,
  Phys.\ Rev.\  D {\bf 78}, 023505 (2008)
  [arXiv:0801.1565 [gr-qc]];
  Z.~K.~Guo, N.~Ohta and S.~Tsujikawa,
  Phys.\ Rev.\  D {\bf 76}, 023508 (2007)
  [arXiv:astro-ph/0702015];
  J.~Q.~Xia,
  Phys.\ Rev.\  D {\bf 80}, 103514 (2009)
  [arXiv:0911.4820 [astro-ph.CO]];
  L.~L.~Honorez, B.~A.~Reid, O.~Mena, L.~Verde and R.~Jimenez,
  JCAP {\bf 1009}, 029 (2010)
  [arXiv:1006.0877 [astro-ph.CO]];
  M.~Baldi,
  arXiv:1005.2188 [astro-ph.CO];
  Y.~Li, J.~Ma, J.~Cui, Z.~Wang and X.~Zhang,
  Sci.\ China\ Phys. Mech. Astron. {\bf 54} (2011) 1367
  [arXiv:1011.6122 [astro-ph.CO]];
  Y.~H.~Li and X.~Zhang,
  Eur.\ Phys.\ J.\ C\ {\bf 71} (2011) 1700
  [arXiv:1103.3185 [astro-ph.CO]].

\bibitem{Suwa}
  M.~Suwa and T.~Nihei,
  Phys.\ Rev.\ D\ {\bf 81} (2010) 023519
  [arXiv:0911.4810 [astro-ph.CO]].

\bibitem{14Amanullah:2010vv}
  R.~Amanullah {\it et al.},
  Astrophys.\ J.\  {\bf 716}, 712 (2010)
  [1004.1711 [astro-ph.CO]].

\bibitem{Nesseris:2005ur}
  S.~Nesseris and L.~Perivolaropoulos,
  Phys.\ Rev.\  D {\bf 72}, 123519 (2005)
  [arXiv:astro-ph/0511040];
  L.~Perivolaropoulos,
  Phys.\ Rev.\  D {\bf 71}, 063503 (2005)
  [arXiv:astro-ph/0412308];
  S.~Nesseris and L.~Perivolaropoulos,
  JCAP {\bf 0702}, 025 (2007)
  [arXiv:astro-ph/0612653].

\bibitem{55Wang:2006ts}
  Y.~Wang and P.~Mukherjee,
  Astrophys.\ J.\  {\bf 650}, 1 (2006)
  [arXiv:astro-ph/0604051].

\bibitem{54Bond:1997wr}
  J.~R.~Bond, G.~Efstathiou and M.~Tegmark,
  Mon.\ Not.\ Roy.\ Astron.\ Soc.\  {\bf 291}, L33 (1997)
  [arXiv:astro-ph/9702100].

\bibitem{57Tegmark:2003ud}
  M.~Tegmark {\it et al.}  [SDSS Collaboration],
  Astrophys.\ J.\  {\bf 606}, 702 (2004)
  [arXiv:astro-ph/0310725];
  M.~Tegmark {\it et al.}  [SDSS Collaboration],
  Phys.\ Rev.\  D {\bf 74}, 123507 (2006)
  [arXiv:astro-ph/0608632].

\bibitem{58Eisenstein:2005su}
  D.~J.~Eisenstein {\it et al.}  [SDSS Collaboration],
  Astrophys.\ J.\  {\bf 633}, 560 (2005)
  [arXiv:astro-ph/0501171].

\bibitem{IC1}
  A.~R.~Liddle,
  Mon.\ Not.\ Roy.\ Astron.\ Soc.\  {\bf 351}, L49 (2004)
  [arXiv:astro-ph/0401198].


\bibitem{IC2}
  M.~Szydlowski and A.~Kurek,
  arXiv:0801.0638 [astro-ph].


\bibitem{IC3}
  T.~M.~Davis, E.~Mortsell, J.~Sollerman, A.~C.~Becker, S.~Blondin, P.~Challis, A.~Clocchiatti and A.~V.~Filippenko {\it et al.},
  Astrophys.\ J.\  {\bf 666}, 716 (2007)  [astro-ph/0701510].


\bibitem{IC4}
  M.~Li, X.~D.~Li and X.~Zhang,
  Sci.\ China Phys.\ Mech.\ Astron.\  {\bf 53}, 1631 (2010)
  [arXiv:0912.3988 [astro-ph.CO]].

\bibitem{IC5}
  S.~del Campo, J.~.C.~Fabris, R.~Herrera and W.~Zimdahl,
  Phys.\ Rev.\ D {\bf 83}, 123006 (2011)  [arXiv:1103.3441 [astro-ph.CO]].

\bibitem{Duran:2010ky}
  I.~Duran and D.~Pavon,
  Phys.\ Rev.\ D {\bf 83}, 023504 (2011)  [arXiv:1012.2986 [astro-ph.CO]].





\end{thebibliography}
\end{document}